\documentclass[aps,pre,preprint, superscriptaddress]{revtex4-2}

\usepackage{graphicx} 
\usepackage{subfigure}
\usepackage[export]{adjustbox} 
\usepackage{amsmath}
\usepackage{float}
\usepackage[T1]{fontenc}
\usepackage[applemac]{inputenc}
\usepackage{xcolor}

\begin{document}

\title{Acoustic transparency and absorption in dense granular suspensions}

\author{Arnaud Tourin}
\email[]{arnaud.tourin@espci.psl.eu}

\affiliation{Institut Langevin, ESPCI Paris, Universit\'e PSL, CNRS, Paris, France}

\author{Yamil Abraham}

\affiliation{Institut Langevin, ESPCI Paris, Universit\'e PSL, CNRS, Paris, France}

\affiliation{Laboratorio de Ac\'ustica Ultrasonora, Instituto de F\'isica, Facultad de Ciencias, Universidad de la Rep\'ublica, Montevideo, Uruguay}

\author{Marie Palla}

\affiliation{Institut Langevin, ESPCI Paris, Universit\'e PSL, CNRS, Paris, France}

\author{Arthur Le Ber}

\affiliation{Institut Langevin, ESPCI Paris, Universit\'e PSL, CNRS, Paris, France}

\author{Romain Pierrat}

\affiliation{Institut Langevin, ESPCI Paris, Universit\'e PSL, CNRS, Paris, France}

\author{Nicolas Benech}

\affiliation{Laboratorio de Ac\'ustica Ultrasonora, Instituto de F\'isica, Facultad de Ciencias, Universidad de la Rep\'ublica, Montevideo, Uruguay}

\author{Carlos Negreira}

\affiliation{Laboratorio de Ac\'ustica Ultrasonora, Instituto de F\'isica, Facultad de Ciencias, Universidad de la Rep\'ublica, Montevideo, Uruguay}
 
\author{Xiaoping Jia}

\affiliation{Institut Langevin, ESPCI Paris, Universit\'e PSL, CNRS, Paris, France}

\date{\today}

\begin{abstract}
We demonstrate the existence of a frequency band exhibiting acoustic transparency in 2D and 3D dense granular suspensions, enabling the transmission of a low-frequency ballistic wave excited by a high-frequency broadband ultrasound pulse. This phenomenon is attributed to spatial correlations in the structural disorder of the medium. To support this interpretation, we use an existing model that incorporates such correlations via the structure factor. Its predictions are shown to agree well with those of the Generalized Coherent Potential Approximation (GCPA) model, which is known to apply at high volume fractions, including the close packing limit, but does not explicitly account for disorder correlation. Within the transparency band, attenuation is found to be dominated by absorption rather than scattering. Measurements of the frequency dependence of the absorption coefficient reveal significant deviations from conventional models, challenging the current understanding of acoustic absorption in dense granular media.
\end{abstract}


\maketitle

\textit{Introduction} - Wave propagation in multiple scattering media has received considerable interest over the past decades. In particular, significant efforts have been devoted to exploiting disorder correlations since the introduction of the concept of \textit{hyperuniformity} by Torquato and Stillinger \cite{PhysRevE.68.041113}. Hyperuniform media are disordered systems characterized by suppressed density fluctuations at large scales, combining aspects of crystalline order with the structural freedom of amorphous materials \cite{TORQUATO20181}. This concept has since inspired extensive research, especially in photonics, where hyperuniform disordered structures have been shown to exhibit remarkable optical properties such as isotropic band gaps and disorder-induced transparency. Notably, Leseur et al. showed that an otherwise opaque medium can be made transparent by arranging scatterers on a hyperuniform point pattern \cite{Leseur:16}. A major research thrust has focused on the design and fabrication of artificial materials that leverage these unconventional wave transport properties turning structural disorder into a functional advantage \cite{doi:10.1073/pnas.0907744106, PhysRevA.88.043822, PhysRevLett.119.208001, Piecchulla2023, Siedentop2024}. While substantial progress has been made in optics in this regard, the investigation of hyperuniformity in acoustics remains at an early stage, with only a few pioneering studies suggesting analogous effects in sound propagation. To date, investigations in acoustics have mostly been restricted to two-dimensional systems involving arrays of scatterers immersed in fluids \cite{PhysRevE.102.053001}.

\begin{figure}
\centering
\subfigure[]{\includegraphics[width=.5 \textwidth]{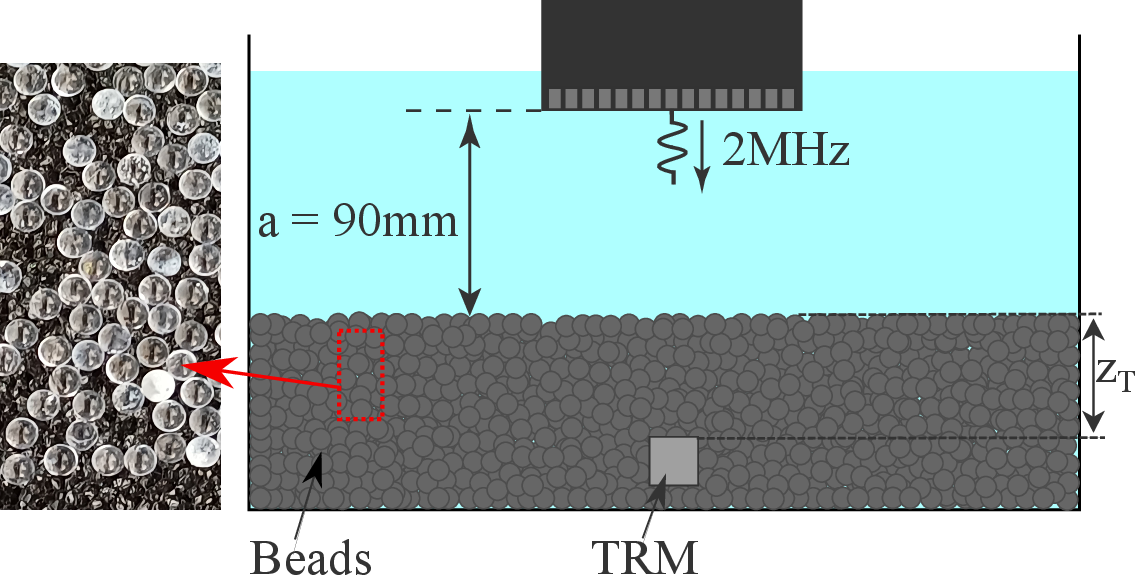}}
\subfigure[]{\includegraphics[width=.5 \textwidth]{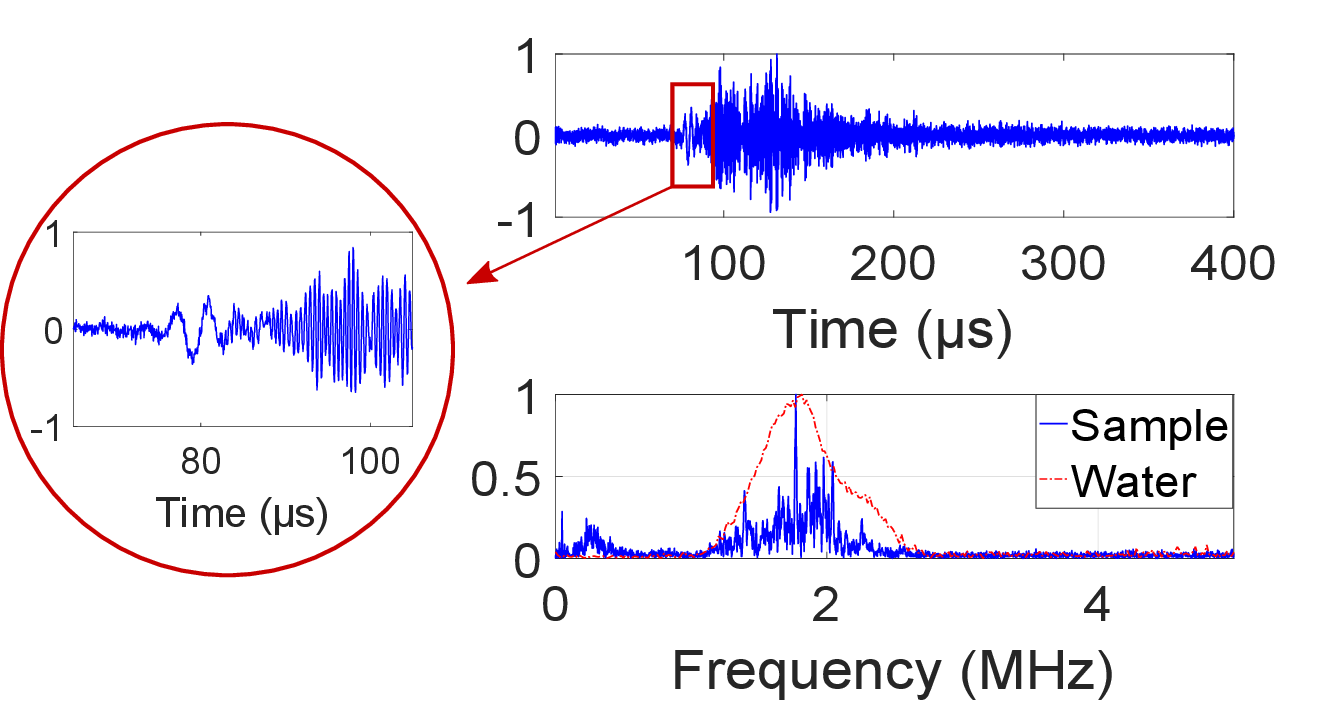}}
\caption{(a) Schematic representation of the experimental setup used for the Time Reversal experiment. The photograph reveals that the beads forming the sample are not perfectly spherical and exhibit slight polydispersity. (b) A typical waveform recorded inside the medium when a two-cycle sinusoidal pulse is emitted by one element of the transducer array. The signal consists of an initial low-frequency arrival - corresponding to the coherent ballistic wave - followed by a long, multiply scattered tail known as the coda. A closer look reveals that the frequencies contained in the ballistic pulse are significantly lower than those in the coda. Comparing the spectrum of the full signal (solid line) to that of a reference pulse transmitted through water (dashed line) highlights how propagation through the suspension enhances the relative contribution of the low-frequency ballistic component. Please note that both spectra are normalized to their respective maxima to facilitate comparison since the amplitude of the signal transmitted through the suspension is much lower than that of the reference.}
\label{fig:figure1}
\end{figure}

Surprisingly, no attempt has been made so far to interpret the acoustic response of compact sphere packings in terms of disorder correlations, even though such configurations were central to the original concept introduced by Torquato and collaborators \cite{PhysRevLett.95.090604}. Our interest in this question emerged unexpectedly during an experiment initially aimed at demonstrating that the diffusion coefficient of ultrasound in a dense granular suspension could be measured using a time-reversal experiment \cite{Abraham2025}. For this purpose, we employed a scattering sample composed of a random close-packed suspension of slightly polydisperse glass beads, with a diameter  $d = 1.5\ \pm \ 0.1\ \mathrm{mm}$, and configured the setup such that the Time Reversal Mirror (TRM) was embedded deep within the scattering medium [see Fig.~\ref{fig:figure1}(a)]. In the first stage of the experiment, a short acoustic pulse was emitted by one element of a transducer array, and the resulting waveform was recorded by a receiver placed inside the suspension. As expected, the recorded signal featured an initial ballistic arrival followed by a long, multiply scattered coda [see the top panel of Fig.~\ref{fig:figure1}(b)].

Unexpectedly, however, a zoom-in on the first part of the signal revealed a striking feature: the central frequency of the ballistic pulse is significantly lower than that of the coda, clearly separating the ballistic and scattered components in both time and frequency. Comparison with the reference signal transmitted through water shows that propagation through the suspension enhances the low-frequency content of the ballistic wave relative to that of the incident pulse [see bottom panel of Fig.~\ref{fig:figure1}(b)]. We verified that the amplitude of this low-frequency ballistic wave scales linearly with the input pulse amplitude. In other words, the medium effectively acts as a linear low-pass filter. As discussed below, this effect arises from strong frequency-dependent scattering attenuation: the low-frequency components propagate essentially ballistically, whereas frequencies near the central frequency of the incident pulse undergo strong scattering.

A somewhat similar behavior was observed by Jia et al. in the late 1990s in dry granular media \cite{PhysRevLett.82.1863}, though the underlying physics is fundamentally different: in dry granular packings, ultrasound propagates through the heterogeneous network of grain contacts. In a dense granular suspension, the acoustic wave travels primarily through the interstitial fluid (water), with weak coupling to the granular skeleton, as evidenced by the measured velocity of the ballistic wave, $c\sim1700$ $\mathrm{m.s^ {-1}}$, closely matching the speed of sound in water. It is consistent with the previous work by Brunet et al. \cite{PhysRevLett.101.138001}: adding a very small amount of liquid to a dry bead assembly leads to the formation of thin liquid films between the beads, resulting in such strong viscous dissipation that wave transmission through the solid skeleton (grain-contact network) is effectively suppressed.

The observation of this same spectral behavior in two physically distinct systems (a percolated solid skeleton in the first case, a pore fluid in the second) suggests that the effect may arise from structural disorder correlations inherent to the bead packing, which is the common feature shared by both media despite the different propagation mechanisms in each case.

\begin{figure}
	\centering 
	\includegraphics[width=.4 \textwidth]{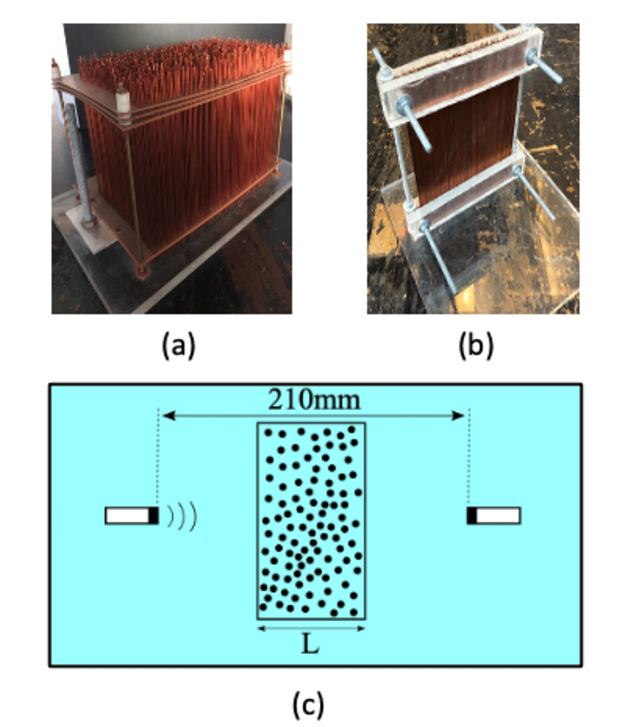}
 	\caption{ Experimental setup (2D). The samples consist of random arrangements of copper cylinders. (a) A dilute random medium with a surface coverage $\Phi_S = 4.4\%$. (b) A compact medium with surface coverage $\Phi_S= 72\%$. (c) The sample is immersed in water between two piezo-electric transducers (diameter: 12.5\, mm) operating at a central frequency of 1\,MHz. The emitting transducer on the left is driven by a three-cycle sine wave at its central frequency.}
	\label{fig:figure2}
\end{figure}

\begin{figure}[htbp]
	\centering 
	\includegraphics[width=0.9 \textwidth]{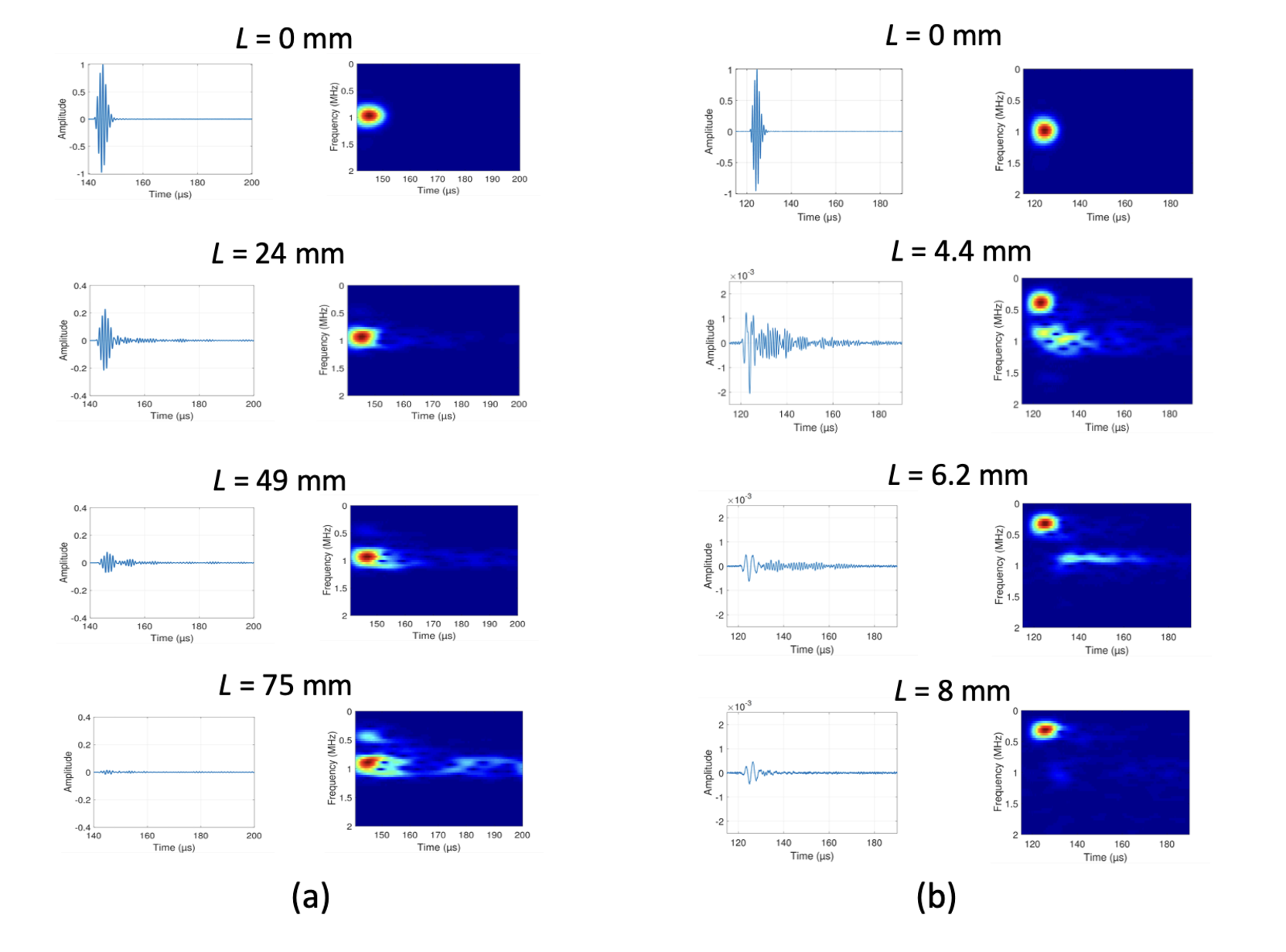}
 	\caption{Averaged transmitted waveforms through a 2D arrangement of copper cylinders and corresponding time-frequency maps, as a function of sample thickness. The wave field is spatially averaged over the surface of the receiving transducer surface, whose diameter is about 2.5 times the wavelength at 0.3 MHz. In panel (a) (respectively, panel (b)), an additional average is taken over 12 (respectively, 10) different relative source-receiver positions with respect to the sample. Note that the number of realizations is not sufficient to fully suppress the coda. Time-frequency analysis is performed using 15$\mu s$-long sliding windows with a step size of 0.3 $\mu s$. (a) Dilute medium: $\Phi_S = 4.4\%$. The coherent wave tends to vanish for the largest sample thickness. (b) Compact medium: $\Phi_S= 72\%$. As the sample thickness increases, high-frequency components tend to disappear, leaving only the low-frequency ballistic wave.}
	\label{fig:figure3}
\end{figure}

\textit{2D cylinder packings} - To test this hypothesis, we began by investigating the acoustic response of two-dimensional scattering samples, which allowed us to more easily explore two limiting cases: a dilute medium, where no effect of disorder correlations is expected, and a densely packed one. As shown in Fig.~\ref{fig:figure2}, the dilute sample consists of a random arrangement of copper cylinders (diameter $d = 0.7\,\mathrm{mm}$) with a surface coverage of $\Phi_S = 4.4\%$. This sample was placed between two transducers operating at a central frequency of 1~MHz. The transmitted waveforms were recorded and then spatially averaged to isolate the coherent wave, whose evolution with sample thickness is illustrated in Fig.~\ref{fig:figure3}(a). For each thickness, the coherent wave was subjected to a time-Fourier transform, and the magnitude of the result was squared  to obtain the frequency-dependent coherent intensity. This intensity decays exponentially with thickness over a characteristic distance known as the extinction length, $\ell_{\mathrm{ext}}$, such that the coherent wave tends to vanish for the largest sample thickness ($L = 75\,\mathrm{mm}$). The extinction length satisfies the relation $1/\ell_{\mathrm{ext}} = 1/\ell_{a} + 1/\ell_{s}$ where $\ell_{a}$ is the absorption mean free path and $\ell_{s}$ the scattering mean free path. At the central frequency (1~MHz), the extinction length was found to be approximately 10\ mm. This result agrees well with predictions from the Independent Scattering Approximation (ISA) for the scattering mean free path \cite{LAGENDIJK1996143}, suggesting that in the dilute regime, attenuation is dominated by scattering, with negligible absorption.

When the dilute medium is replaced by a random close packing of the same copper cylinders, a low-frequency ballistic pulse followed by a higher-frequency coda wave is observed, similar to the behavior seen in the 3D case [see left column of Fig.~\ref{fig:figure3}(b)]. As the sample thickness increases, high-frequency components tend to disappear, leaving only the low-frequency ballistic wave. As shown in the right column of Fig.~\ref{fig:figure3}(b), this progressive loss of high-frequencies results in a downward shift of the transmitted spectrum to about 0.3\,MHz, bringing it closer to the lower edge of the bandwidth of the incident pulse. The remaining spectral components lie within a frequency range where scattering is significantly reduced, allowing the wave to propagate in an almost purely ballistic manner. This frequency range can therefore be referred to as a \textit{transparency window}. This transparency window extends up to about 0.5\,MHz, corresponding to $kd \sim 1.5$, where $k$ is the wavenumber in water. This lies well beyond the homogenization regime \cite{sheng2006introduction}, characterized by $ks\ll 1$, where $s$ denotes the average spacing between cylinders, which is here is of the order of the cylinder diameter $d$; in that regime, transparency would be trivially expected. In other words, although the wave undergoes scattering, the resulting wave field fluctuations tend to cancel out due to disorder correlations.  As in the 3D case, this suggests that disorder correlations must be taken into account to explain the observed behavior. The extinction length measured is 2.1\,mm at 0.3\,MHz , i.e., roughly three cylinder diameters, a surprisingly short distance for a frequency range described as a transparency window. However, this strong attenuation is likely due to absorption rather than scattering, and therefore does not contradict the notion of transparency in the sense of minimal wave scattering. We will return to this point later in the discussion.

\textit{3D sphere packings} - It has long been known that structural correlations can reduce (or enhance) scattering efficiency. This is, for example, the physical basis for the transparency of the cornea to visible light \cite{Hart:69}. Such reductions in scattering efficiency can be quantified using an approach independently proposed by Saulnier \cite{PhysRevB.42.2621} and Maret \cite{PhysRevLett.65.512}. Given the form factor $F(q)$ of individual scatterers, which describes their response to an incident field as a function of the scattering wave number $q$, and the structure factor, defined as $S(q) =\frac{1}{N} \left | \sum_{j=1}^{N} e^{+i \textbf {q}\cdot \textbf{r}_j} \right |^2$, with $N$ the number of scatterers and $\textbf{r}_i$ their positions, one can compute a corrected, or effective, scattering cross section, that incorporates interference effects between waves scattered by different scatterers:

\begin{equation}
\sigma_{sc}^{SF} = \frac{2\pi}{k_r^4} \int_0^{2k_r} F(q) S(q) q dq
\label{eq:eq1} 
\end{equation}
with $k_r$ the real part of the effective wave number \cite{RevModPhys.95.045003}.

\begin{figure}[htbp]
\centering
\subfigure[$ \Phi_V = 0.4$]{\includegraphics[width=.45 \textwidth]{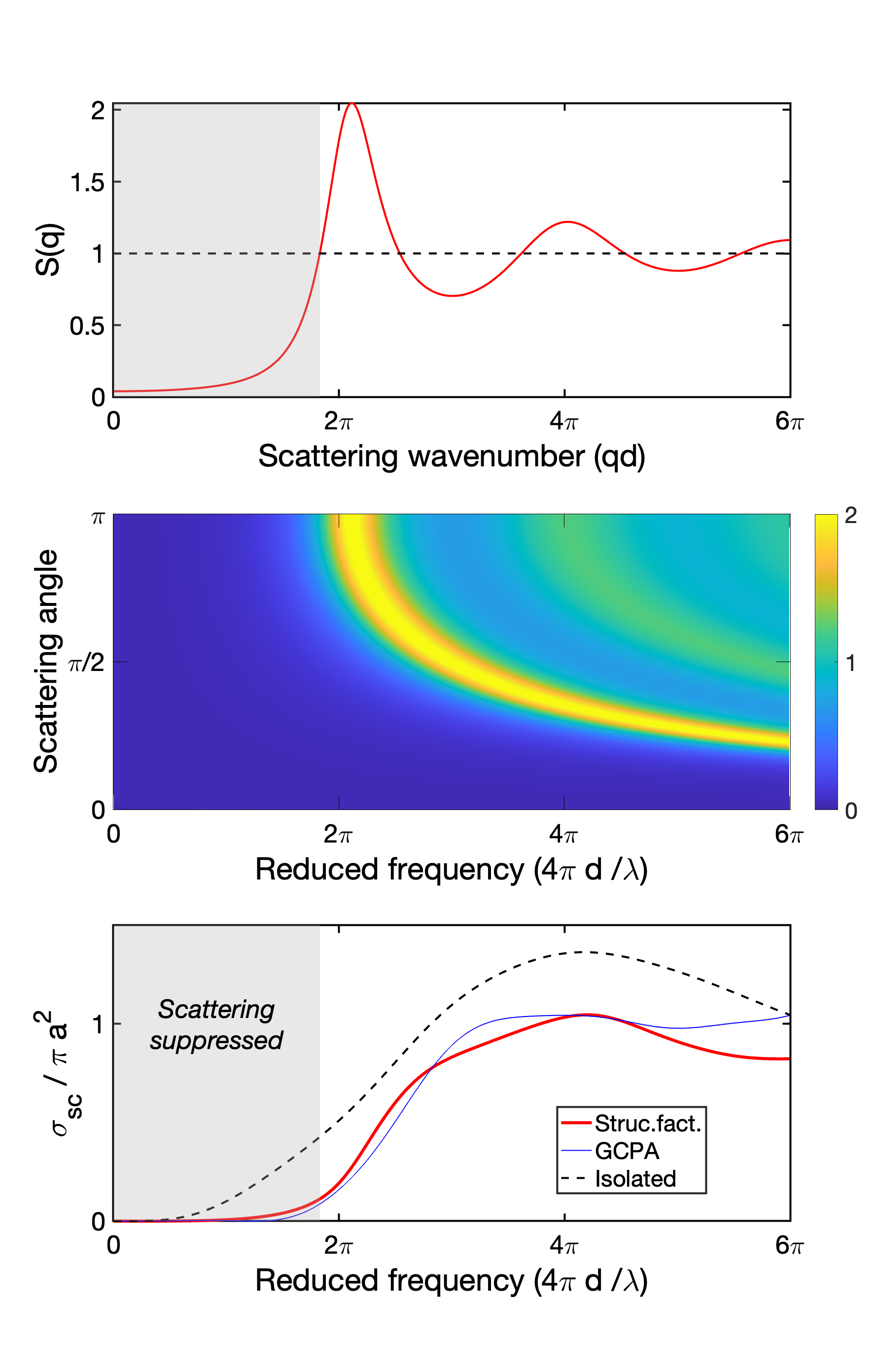}}
\subfigure[$\Phi_V = 0.64$]{\includegraphics[width=.45 \textwidth]{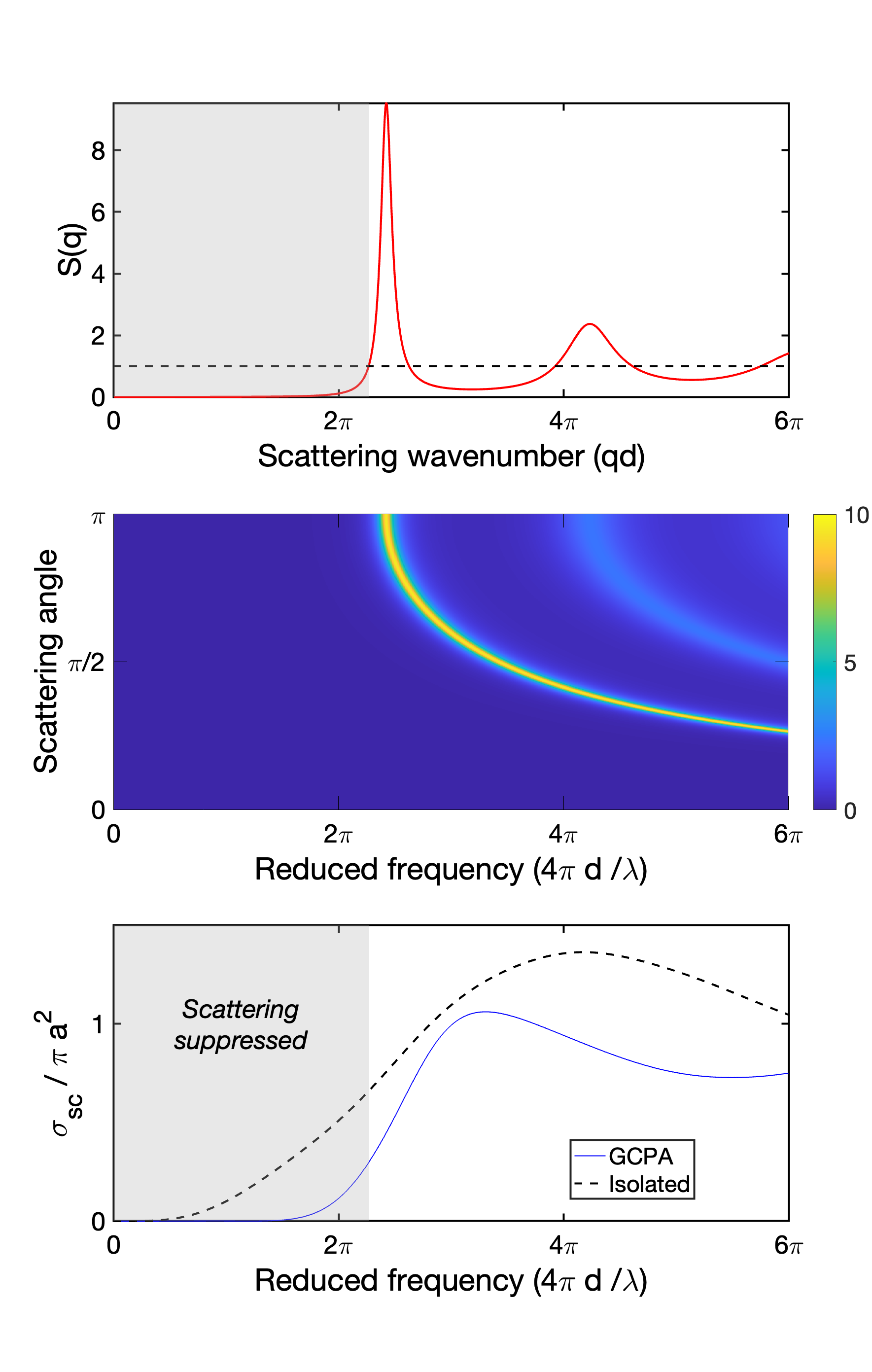}}
\caption{Impact of structural correlations on ultrasound scattering in dense granular suspensions, for two packing fractions of glass beads. Top: structure factor $S(q)$, derived analytically using the Percus - Yevick  approximation, as a function of the reduced scattering wave number $qd$, with $d$ the bead diameter. The gray-shaded region highlights the low-$q$ range where $S(q)$ falls below unity (except at $q=0$ where $S(0)=1$ in any case). Middle: structure factor plotted as a function of scattering angle and frequency, using $q=\frac{4\pi}{\lambda}\sin(\theta/2)$, where $\lambda$ is the wavelength in water and $\theta$ is the scattering angle. Bottom: effective scattering cross section. Predictions based on the structure factor $S(q)$ (thick solid red line) and the GCPA model (thin solid blue line) show good agreement. In particular, both models predict a vanishing cross section within the frequency band marked by the gray-shaded area. For reference, the theoretical scattering cross section of a single glass bead \cite{10.1121/1.1906780} is shown as a dotted line.}
\label{fig:figure4}
\end{figure}

Let us now evaluate the ability of this theoretical approach to account for the transparency window observed in the 3D glass bead packing.  As a first step, we consider a suspension of glass beads (diameter $d=1.5\mathrm{mm}$) in water with a volume fraction $\Phi_V = 40\%$. Although this specific case is not experimentally studied here, it is of particular interest because the structure factor can be derived analytically using the Percus - Yevick (PY) approximation \cite{PhysRevLett.10.321}. The PY approximation applies to dense assemblies of monodisperse hard spheres,  i.e., rigid particles with purely repulsive, short-range interactions. The resulting structure factor is shown in Fig. \ref{fig:figure4}(a), either as a function of the scattering wavenumber (top panel), or as a function of both scattering angle $\theta$ and frequency (middle panel) using $q=\frac{4\pi}{\lambda}\sin(\theta/2)$ with $\lambda$ the wavelength in water. Two main features emerge from the middle panel : first, the structure factor vanishes for all frequencies in the vicinity of the forward direction $\theta = 0$ (but not exactly along it where it is equal to unity by definition), revealing a strong reduction in scattering thanks to Eq. \ref{eq:eq1}. Second, in the backward direction $\theta = \pi$, the structure factor reaches a maximum at frequencies around $qd = 2\pi$, i.e., corresponding to a wavelength twice the average distance (i.e., two bead diameters) between beads. This peak has an important contribution for the computation of the scattering cross-section, meaning that there is a strong backscattering. This is reminiscent of the Bragg condition in ordered system.

Using this structure factor in Eq.~(\ref{eq:eq1}), we can compute the effective scattering cross section. Note that, instead of using $k_r$, we use $k$, the wave number in water, as the upper bound of the integral. This approximation introduces only a small difference since, over the explored frequency range, the coherent wave speed is close to that in water (approximately 1700 $\mathrm{m s^ {-1}}$) with minimal dispersion. As seen in the bottom panel of Fig.~\ref{fig:figure4}(a), this scattering cross section vanishes in the low-frequency range where $\lambda > d/2$  (gray-shaded area), revealing the existence of a transparency window in which wave propagation remains almost purely ballistic. This window coincides with the range where the structure factor falls below unity. We also plot the prediction from the Generalized Coherent Potential Approximation (GCPA) \cite{PhysRevLett.66.1240} that provides an alternative, albeit less explicit, way to account for disorder correlations. A good agreement is obtained. Let us now consider the case of a close-packed configuration, with a volume fraction of $\Phi_V = 64\%$ as in our experiment. The analytical structure factor derived from the PY approximation still captures the position of the first peak reasonably well, but it overestimates its amplitude [Fig. \ref{fig:figure4}(b)], as evidenced by comparison with the structure factor obtained from a simulated jammed packing of $10^6$ particles \cite{PhysRevLett.95.090604}. This discrepancy is expected, as the PY approximation is accurate over a broad range of densities within the liquid phase, but it breaks down at volume fractions approaching the close-packing limit \cite{PhysRevE.99.012146}. To compute the effective scattering cross section, we therefore rely on the GCPA only. The frequency range where the effective scattering cross section vanishes [solid line in the bottom panel of Fig. \ref{fig:figure4}(b)] extends up to approximatively $\frac{4\pi d}{\lambda} \sim 2 \pi$ (i.e., $kd \sim \pi$), which again closely matches the interval where the structure factor drops below 1 (gray-shaded area).

Building on the theoretical predictions discussed above, we now turn to a new set of acoustic transmission experiments to verify whether the predicted transparency window appears at the expected frequencies.
\begin{figure}[htbp]
	\centering 
	\subfigure[]{\includegraphics[width=.4 \textwidth, valign=c]{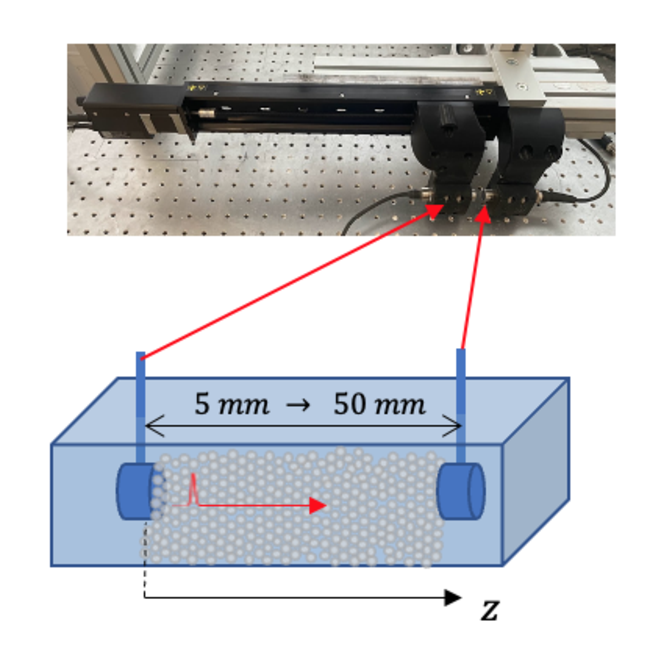}}
	\subfigure[]{\includegraphics[width=.4 \textwidth, valign=c]{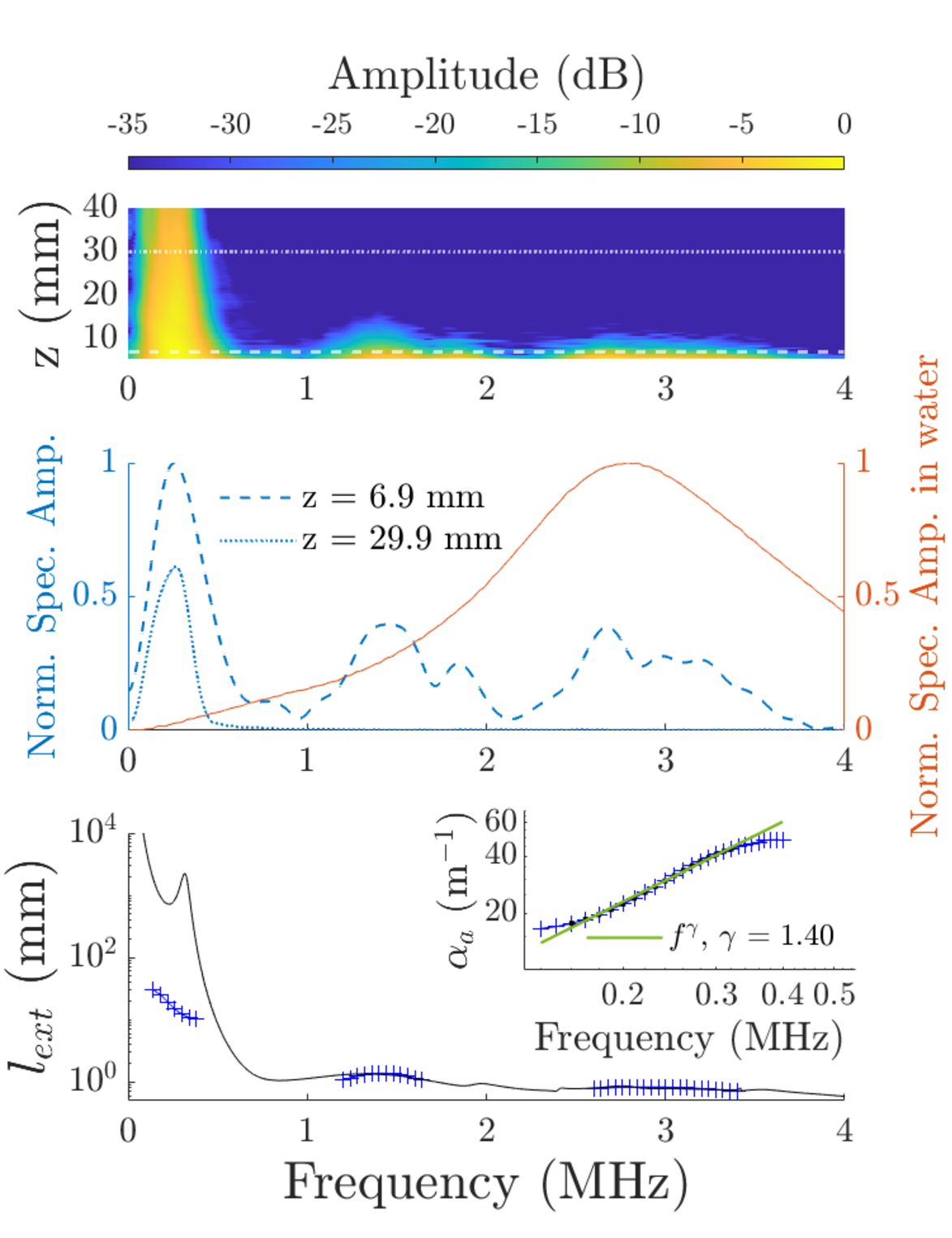}}
 	\caption{(a) Experimental setup used to measure acoustic transmission through a dense granular suspension of glass beads (diameter $d=1.5\mathrm{mm}$). The emitting transducer is excited with a half-period sine wave at its central frequency (3.5 MHz). The receiving transducer is mounted on a computer-controlled translation stage, allowing signal acquisition over source-receiver distances ranging from 5 mm to 50 mm in  steps of 0.5 mm. (b) Top: transmitted signal amplitude as a function of frequency and propagation distance. Middle: spectra extracted from the top panel at two propagation distances and comparison with the spectrum in water (red solid line). The three spectra are normalized to their respective maxima. Bottom: measured extinction length (crosses) compared with the GCPA prediction for the scattering mean free path (solid line). In the low frequency range, the corresponding extinction coefficient follows a power law (see inset).}
	\label{fig:figure5}
\end{figure}
The experimental setup is illustrated in Fig.~\ref{fig:figure5}(a). The sample was immersed in water between two identical ultrasonic transducers operating at a central frequency of 3.5 MHz. One transducer served as the emitter, driven by a half-period sine wave at its central frequency. The receiving transducer was mounted on a computer-controlled translation stage, allowing transmission measurements over a wide range of source-receiver distances (from 5 mm to 50 mm, with a step size of 0.5 mm). Fig.~\ref{fig:figure5}(b) (top and middle panels) shows how the transmitted signals evolve as a function of frequency and propagation distance. Remarkably, although the transducers are centered at a relatively high frequency, the combination of the broadband excitation and the very low attenuation within the transparency window allows the low-frequency components to dominate the transmitted signal. This makes it possible to clearly observe the transparency window predicted by the model, which extends up to $\sim$ 0.5 MHz (i.e., $kd \sim \pi$) in good agreement with the frequency range where the effective scattering cross section is predicted to vanish. Two other frequency ranges with reduced scattering are observed around 1.4 MHz and 3 MHz, corresponding to local maxima in the extinction length, with the first one consistent with the second dip observed at $d \sim 1.4 \lambda$  in the scattering cross section shown in the bottom panel of Fig.~\ref{fig:figure4}(b).

To quantify attenuation, we followed the same procedure as in the 2D experiments to extract the extinction length $\ell_{\text{ext}}$ as a function of frequency, plotted as blue crosses in the bottom panel of Fig.~\ref{fig:figure5}(b). The extinction length is found to decrease from 30 mm to 10 mm within the transparency band. Experimental data show very good agreement with GCPA predictions at frequencies above 1~MHz [solid line in the bottom panel of Fig.~\ref{fig:figure5}(b)]. In contrast, at lower frequencies (within the transparency band where the predicted scattering mean free path can exceed one hundred bead diameters), experimental data deviate significantly from the GCPA predictions. Note that in their seminal work, Page et al. \cite{page1995experimental} did not report extinction length measurements for $kd < 4$. It is important to emphasize that the GCPA model accounts only for scattering-induced attenuation and does not include absorption. We therefore interpret the observed discrepancy as a signature that, in this frequency range, scattering is sufficiently weak for absorption to become the dominant attenuation mechanism. As a result, the measured extinction length corresponds to the absorption length. 

In the transparency window, the frequency dependence of the corresponding absorption coefficient, defined as $\alpha_{\text{a}}=\frac{1}{2\ell_{\text{a}}}$, follows a power law: $\alpha_a \sim f^{\gamma}$, with $\gamma \approx 1.4$ [see the inset in the bottom panel of Fig.~\ref{fig:figure5}(b)]. This behavior departs significantly from standard theoretical models, which predict $\gamma = 0.5$ for viscous dissipation localized in thin boundary layers between glass beads and the pore liquid (water). For example, the model, as described in \cite{Weaver1995}, predicts that the absorption coefficient is given by $\alpha_a = \Phi_V /(1-\Phi_V) d^{-1} c^{-1} \sqrt{4 \pi \mu f/ \rho}$, with $\mu$ and $\rho$ the viscosity and mass density of water, respectively. The absorption length estimated from this model gives $\ell_{\text{a}} = 400$ mm at 0.3 MHz corresponding to $260$ bead diameters.  This value is more than one order of magnitude larger than the measured value $\ell_{\text{ext}} \sim \ell_{\text{a}} \approx 20 ~\text{mm}$. The measured values also contrast with earlier experimental studies, where exponents typically lie in the range $0.5 < \gamma < 1$ \cite{Salin1981, Weaver1995}.

\textit{Concluding remarks} - Our main finding is the identification of an acoustic transparency window in dense granular suspensions, allowing the propagation of a low-frequency ballistic wave excited by a broadband pulse emitted from a high-frequency transducer. Although such experimental packings of slightly polydisperse glass beads immersed in water are unlikely to satisfy the conditions required for a hyperuniform medium, they clearly exhibit at least short-range correlations. We argued that the observed transparency window can be interpreted as a consequence of these correlations. To support this interpretation, we used an existing model that incorporates such correlations via the structure factor. We showed that its predictions agree well with those of the Generalized Coherent Potential Approximation (GCPA) model, which applies at high volume fractions, including the close-packing limit, but does not explicitly include disorder correlations. In the frequency range of the transparency window, unexplored in the seminal work by Page et al. \cite{page1995experimental}, attenuation appears to be dominated by absorption rather than scattering so that the GCPA model, which does not incorporate absorption, fails to fit the experimental data. Moreover, the observed frequency dependence of absorption does not follow existing theoretical predictions, nor does it align with previous experimental results. These results motivate further experimental investigation to better understand the mechanisms driving attenuation in such systems. To this end, we plan to expand our study by exploring a wider range of volume fractions, as well as varying bead materials and sizes.

\begin{acknowledgments}
\textit{Acknowledgments}-This work was supported by the French government under the program ``Investissements d'Avenir'', by the Agencia Nacional de Investigaci\'on e Innovaci\'on (ANII), Uruguay, and by the Institut Franco-Uruguayen de Physique (LIA IFUP) 
\end{acknowledgments}

\end{document}